\documentclass[12pt,a4paper]{article}
\usepackage{graphicx,amsmath}
\usepackage[mathscr]{eucal}
\usepackage{psfrag}
\newcommand{\eps}{\varepsilon}
\renewcommand{\vec}[1]{\mathbf{#1}}
\renewcommand{\Re}{\textrm{Re}\,}

\begin{document}

\title{Scattering of evanescent wave by two cylinders\\ near a flat boundary}

\author{Oleg V. Belai$^1$, Leonid L. Frumin$^{1,2}$, Sergey V. Perminov$^3$,  \\David A. Shapiro$^{1,2}$\\ \ \\
$^1$Institute of Automation and Electrometry,\\ Siberian Branch, Russian
Academy of Sciences,\\ 1 Koptjug Avenue, Novosibirsk, 630090 Russia\\
$^2$Novosibirsk State University,\\ 2 Pirogov Street, Novosibirsk
630090, Russia\\
$^3$A.V. Rzhanov Institute of Semiconductor Physics,\\ Siberian Branch,
Russian Academy of Sciences,\\ 13 Lavrentjev Avenue, Novosibirsk 630090, Russia}

\maketitle

\begin{abstract}
Two-dimensional problem of evanescent wave scattering by dielectric
or metallic cylinders near the interface between two dielectric
media is solved numerically by boundary integral equations method. A special Green function was proposed to avoid the
infinite integration. A pattern with a circular and a prolate elliptic cylinders, respectively, is suggested to simulate the sample and the probe in near-field optical microscopy.   The energy flux in the midplane of the probe-cylinder is
calculated as a function of its position.

\end{abstract}

The diffraction limit in optics, known to originate from the wave nature of light, gives a striking example of a physical restriction being a target of increasing efforts to overcome. More than a century ago it was
realized that the wavelength limits the smallest spot within the
electromagnetic energy can be localized, as well as the smallest
details one can optically resolve are comparable to the wavelength. However,
further study showed that these obstacles, in fact, concern a traveling
(homogeneous) electromagnetic wave.  Unlike, the
inhomogeneous (also referred to as {\em evanescent}) waves, which can not
propagate far away from their source, open the way to clear up the limitations
due to diffraction and go towards the optics of tiny objects. For instance, the
nanosized highly-polarizable (i.e. metal) particles well manage to concentrate
the light energy within few nanometer range \cite{Stockman11}.  A {\em
near-field scanning optical microscopy} (NSOM) was suggested to obtain optical
signal from the objects at nanoscale (see \cite{Greffet97,Hecht00} and references
therein) using sharp tips; the latter serve much like an optical antenna
\cite{Navotny11}, which receives the energy of the local field and then
transfers it to a detector. Thus, the nanophotonics is, basically, an
optics of evanescent waves, and, consequently, the fundamental optical
processes (such as diffraction, interference, scattering) are to be reconsidered.

In past two decades a substantial progress is achieved in nano-optics
\cite{Novotny_book,Girard05}. However, a significant methodological deficiency persists
even for the plain, basic problems, like scattering of the evanescent wave by a
body. The trouble is that evanescent wave can not be considered in isolation
from its source (for instance, the interface where the total internal reflection
takes place), therefore the source is certainly affected by the scatterer as being located within a few
wavelengths. In paper \cite{olBFPS11} the general analytical approach
is suggested that makes possible to do very effective calculations of the
evanescent wave scattering on a 2D particle (a cylinder) near a flat boundary.
In the present work we make the next step and consider the problem of two
optically coupled objects placed into the inhomogeneous wave. Our main goal is
to get a physical insight into the near-field scanning optical microscopy,
which minimally involves two small bodies --- the studied object and the probe.
We believe our work is a useful starting point for analysis
of particular NSOM schemes which will allow for correct extraction of the near field and structural information from NSOM data. Keeping in mind this application to the realistic configurations, we should focus our attention on
the first-principles approaches, avoiding restricting assumptions and approximations.

We start from the Helmholtz equation
\begin{equation}\label{Helmholtz}
    (\bigtriangleup+k^2)\mathscr{H}=0,
\end{equation}
where  $\bigtriangleup$ is the Laplace operator, $\vec{k}$  is the wavevector.
Consider domain $\mathscr{D}$ with permittivity $\eps_{\textrm{in}}$ and its
boundary $\Gamma=\partial\mathscr{D}$.  Denote $\eps_{\textrm{out}}$ the
permittivity of exterior of $\mathscr{D}$. The Green theorem can be written inside and outside domain $\mathcal{D}$, respectively
\begin{eqnarray}\label{Green-theorem-in}
  \mathscr{H}(\vec{r}) =-\oint_\Gamma
  \left(
  \frac{\partial g}{\partial n'}\mathscr{H}' -g\frac{\partial \mathscr{H}'}{\partial n'}
  \right)\,ds',\\
  \mathscr{H}(\vec{r}) =\oint_\Gamma
  \left(
  \frac{\partial g}{\partial n'}\mathscr{H}' -g\frac{\partial \mathscr{H}'}{\partial n'}
  \right)\,ds'+\mathscr{H}_0(\vec{r}).
  \label{Green-theorem-out}
\end{eqnarray}
Here $\vec{r}\notin\Gamma$, $\mathscr{H}'\equiv\mathscr{H}(\vec{r}')$,
$\partial/\partial n'$ is the derivative along the internal normal,  $g(\vec{r},\vec{r}\,')$ is a fundamental
solution to the inhomogeneous Helmholtz equation
\begin{equation}\label{inhom-equation}
(\bigtriangleup+k^2)g=\delta(x-x')\delta(y-y').
\end{equation}
It should be noted that function $g(\vec{r},\vec{r}\,')$ is not
fully arbitrary, namely, should satisfy the radiation condition. Hence,
the implicit integral over an infinitely remote contour in Eq.
\ref{Green-theorem-out} reduces to the field in the absence of
scatterer $\mathscr{D}$, denoted as $\mathscr{H}_0$.

Let us consider TM wave for which $\mathscr{H}$ is the magnetic field. It has
to satisfy boundary conditions for the field and its normal derivative
\begin{equation}\label{Boundary}
\left[\mathscr{H}\right]_\Gamma= \left[\frac1\eps\frac{\partial \mathscr{H}}{\partial
n}\right]_\Gamma=0,
\end{equation}
where square brackets denote the jump,
$\eps$ corresponds to either $\eps_{\rm in}$ or $\eps_{\rm out}$. Conditions
(\ref{Boundary}) mean that the magnetic field is always continuous
at $\Gamma$, whereas its normal derivative has a jump depending on $\eps_{\rm
in}$, $\eps_{\rm out}$.

In order to find the field with the help of the Green theorem
(\ref{Green-theorem-in}), (\ref{Green-theorem-out}) we need to know
$\mathscr{H}$ and $\partial \mathscr{H} / \partial n$ at boundary $\Gamma$.
These two \emph{independent} functions satisfy the following coupled equations
at $\vec{r} \in \Gamma$:
\begin{eqnarray}
  \frac12\mathscr{H}(\vec{r}) =-\oint_\Gamma
  \left(
  \frac{\partial g_{\textrm{in}}}{\partial n'}\mathscr{H}' -g_{\textrm{in}}\frac{\eps_{\textrm{in}}}{\eps_{\textrm{out}}}\frac{\partial \mathscr{H}'}{\partial n'}
  \right)\,ds',\label{BEM1}\\
\frac12\mathscr{H}(\vec{r}) =\oint_\Gamma
  \left(
  \frac{\partial g_{\textrm{out}}}{\partial n'}\mathscr{H}' -g_{\textrm{out}}\frac{\partial \mathscr{H}'}{\partial n'}
  \right)\,ds'+\mathscr{H}_0(\vec{r}),\label{BEM2}
\end{eqnarray}
which are obtained by approaching $\Gamma$ from either inner or outer domains.
Here the fundamental solution inside and outside $\mathscr{D}$ is denoted as
$g_{\textrm{in}}$ and $g_{\textrm{out}}$, respectively.

After solving coupled integral equations (\ref{BEM1}), (\ref{BEM2}) the field
in arbitrary point can be calculated using (\ref{Green-theorem-in}),
(\ref{Green-theorem-out}). This approach is the base for
the boundary element method (BEM) \cite{Colton,Brebbia84}. Its advantage
consists in diminishing the problem dimension. For instance, in two-dimensional
geometry the method deals with one-dimensional contour $\Gamma$, and then
occurs to be very fast and accurate. The method could be extended to several
domains. For this purpose we must consider integration (\ref{BEM1}),
(\ref{BEM2}) along an unlinked manifold $\Gamma$ with corresponding fundamental solution $g_{\textrm{in}}$ inside each.  External function  $g_{\textrm{out}}$
should satisfy the Sommerfeld radiation condition at infinity, i.e. be the
diverging spherical or cylindrical wave.

\begin{figure}\centering
\includegraphics[width=0.5\columnwidth]{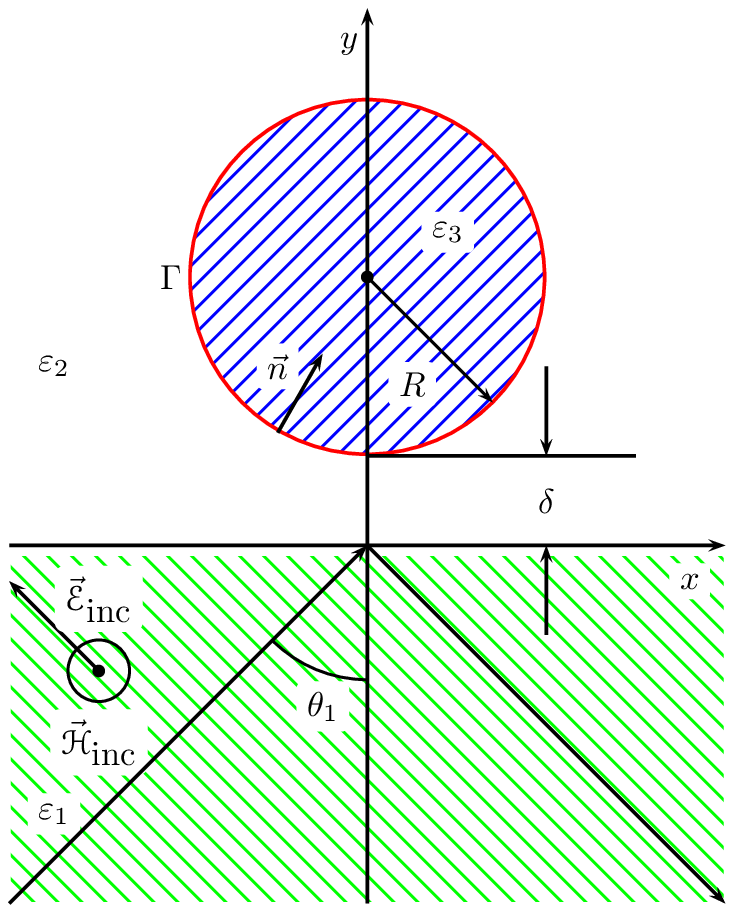}
\caption{The geometry of light scattering.}\label{fig:geometry}
\end{figure}

We consider an evanescent wave, Fig.~\ref{fig:geometry}. The plane running wave
$\mbox{\mathversion{bold}$\mathscr{H}$}(\vec{r},t)=
\mbox{\mathversion{bold}$\mathscr{H}$}_{\textrm{inc}} \exp\left(-i\omega
t+i\vec{k}_1\cdot\vec{r}\right)$ goes from the dielectric media $\eps_1$ to
other medium with permittivity $\eps_2$; $\theta_1$ is the incident angle
between the wavevector $\vec{k}_1$ and the normal to the boundary. While it is
greater than $\theta_{0}=\arcsin\sqrt{{\eps_2}/{\eps_1}}$, the angle of the
total internal reflection, only the evanescent wave with coordinate dependence
$\exp(-\kappa y+ik_{2x}x)$ penetrates into medium 2, where
\begin{equation}\label{decrement}
\kappa=\frac{\omega}c\sqrt{\eps_1\sin^2\theta_1-\eps_2},\quad
k_{2x}=k_{1x}=\frac{\omega}c\sqrt{\eps_1}\sin\theta_1,
\end{equation}
$\omega,c$ are the frequency and speed of light.

Two independent polarization states of falling wave are possible. We consider
TM-wave with magnetic field vector perpendicular to the plane of incidence.
This case is more interesting in view of the plasmon resonances study, since
electric field vector lies in $xy$ plane where the cylinder has a finite size.
The solution for TE-wave can be considered in the same way. The magnetic field
of the wave obeys the Helmholtz equation (\ref{Helmholtz}), then the BEM method
is applicable. However, the problem arises with the infinite integration path
along $x$-axis that is hard for numerical calculation.

To avoid this difficulty we look for the specific Green function $G(x,y;x',y')$
satisfying inhomogeneous equation (\ref{inhom-equation}) in media 1 and 2.
Function $G$ depends on difference $x-x'$ only, due to translational symmetry.
The boundary condition at $y=0$ is
\begin{equation}\label{BoundaryG}
\left[G(x,y;x',y')\right]_{y=0}= \left[\frac1\eps\frac{\partial
G(x,y;x',y')}{\partial y}\right]_{y=0}=0.
\end{equation}
After the Fourier transformation
\begin{equation}\label{Fourier}
G(x,y;0,y')=\frac1{2\pi}\int_{-\infty}^\infty
G_q(y,y')e^{iqx}\,{dq}
\end{equation}
Eq. (\ref{inhom-equation}) is reduced to an ordinary differential equation
having exponential solutions. Using conditions (\ref{BoundaryG}) we obtain
the function at $y'>0$ in $q$-domain
\begin{equation}\label{Green-Fourier}
G_q=-\frac1{2\mu_2}\begin{cases}
\left[1+r(q)\right]e^{\mu_1 y-\mu_2 y'},&y<0,\\
e^{-\mu_2|y-y'|}+r(q)e^{-\mu_2|y+y'|},
& y>0.
\end{cases}
\end{equation}
Here $ r(q)=(\eps_1\mu_2-\eps_2\mu_1)/(\eps_1\mu_2+\eps_2\mu_1)$ is the
Fresnel reflection coefficient of $p$-wave at normal incidence \cite{R09},
$\mu_{1,2}^2=q^2-k_{1,2}^2$. Carrying out Fourier transformation
(\ref{Fourier}) of function (\ref{Green-Fourier}) at $y>0$ we have two terms
\begin{eqnarray}\label{first-Green}
G_1=-\int\limits_{-\infty+i0}^{\infty-i0}
e^{-\mu_2|y-y'|+iq(x-x')}\frac{dq}{4\pi\mu_2},\\
G_2=-\int\limits_{-\infty+i0}^{\infty-i0}
e^{-\mu_2|y+y'|+iq(x-x')}\frac{r(q)\,dq}{4\pi\mu_2}.
\label{second-Green}
\end{eqnarray}
and $G=G_1+G_2$, where the sign of square root is given by the rule
$\sqrt{q^2-k_2^2}\to-i\sqrt{k_2^2-q^2}, q^2<k_2^2$.

The first term (\ref{first-Green}) can be calculated analytically and reduces
to the Green function in the homogeneous space
\begin{equation}\label{Fundamental}
G_1(\vec{r};\vec{r}\,')=\frac1{4i}{H_0^{(1)}
\left(k_2\rho_-\right)},
\end{equation}
where $H_0^{(1)}$ denotes the Hankel function of the first kind \cite{Olver10},
$\rho_{\pm}^2=(x-x')^2+(y\pm y')^2$. The second term $G_2$ gives the effect of
the reflected image source. The amplitude of source at each $q$ is equal to the
reflection coefficient $r(q)$. Thus along with the point source at $(x',y')$ we
have to consider the mirror-image source $r(q)$ at $(x',-y')$. The total field
at the upper half-plane is the sum of fields generated by the source and its
image at each $q$. The Green function automatically takes into account the multiple
scattering. The Green function of this type was studied for
homogeneous waves: spherical acoustic, see  \cite{B80,LL6}, or cylindrical
electromagnetic waves \cite{Pincemin94}.

The asymptotic behavior of the Green function in far field can be found by the
steepest descent method \cite{BH86}. The stationary point is
$q_0=k_2|x-x'|/\rho$, where $\rho=\rho_-$ for $G_1$ and $\rho=\rho_+$ for
$G_2$. The result is a sum of cylindrical waves $G\propto
{\rho_-^{-1/2}}{e^{ik_2\rho_-}}+r_0 {\rho_+^{-1/2}}{e^{ik_2\rho_+}},$ where
\begin{equation}\label{image}
r_0=
\frac%
{\eps_1\sin \varphi-\sqrt{\eps_2(\eps_1-\eps_2\cos ^2\varphi)}}
{\eps_1\sin \varphi+\sqrt{\eps_2(\eps_1-\eps_2\cos ^2\varphi)}}
\end{equation}
is the reflection coefficient at $q=q_0$,
$\varphi$ is the polar angle of observation counted out from $x$-direction. The
reflection coefficient is $r_0=-1$ at $\varphi=0$ and
$(\sqrt{\eps_1}-\sqrt{\eps_2})/(\sqrt{\eps_1}+\sqrt{\eps_2})$ at
$\varphi=\pi/2$. It turns to zero at $\varphi=\arctan\sqrt{\eps_1/\eps_2}$,
i.e. at the Brewster angle.

Green function (\ref{first-Green}), (\ref{second-Green}), exploited as the
external fundamental solution for coupled integral equations (\ref{BEM1}),
(\ref{BEM2}), allows us to avoid the infinite integration along axis $x$. Then
we can solve the equations for two contours. The algorithm has been tested in
the case of one contour and homogeneous wave when $\eps_1=\eps_2$. There are
analytical formula for circular cylinder \cite{Harrington61} and numerical
calculations for a cylinder with elliptic cross-section \cite{jtpGZ06}.
Comparison demonstrates the relative consistency within $10^{-4}$ for $N=360$ panels
approximating contour $\Gamma$.

\begin{figure}\centering
\includegraphics[width=0.9\columnwidth]{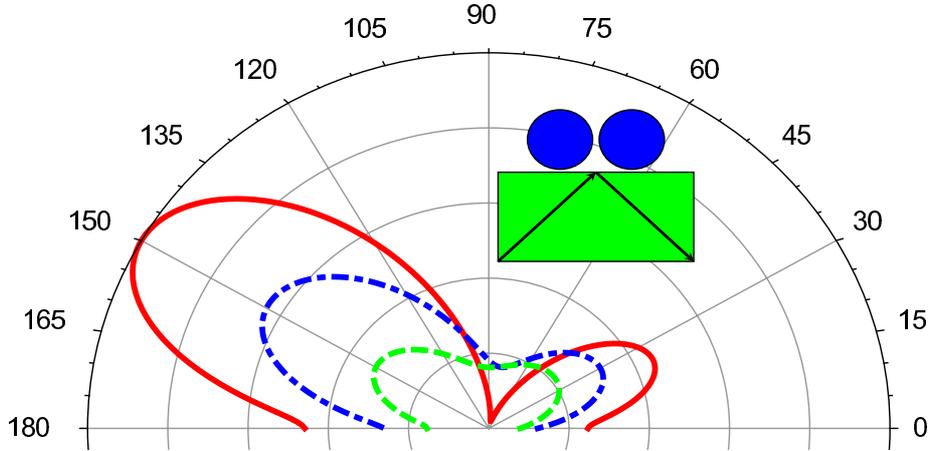}
\caption{Energy flux through a distant semicircle (arb. units) from two equal
circles with distance between centers $L=0.21~\mu$m as a function of polar
angle $0<\varphi<\pi$ at $\lambda=1.512~\mu$m, $R=0.1~\mu$m,
$\delta=0.01~\mu$m, $\theta_1=\pi/4, \eps_2=1, \eps_3=\eps_4=-91.5+10.3i$, and
different  $\eps_1=2$ (solid line), 2.25 (dot-dahed), 3
(dashed).}\label{fig:double}
\end{figure}

Without scattering body the electric field vector
{\mathversion{bold}${\mathscr{E}}$} in medium 2 has only $y$-component.
The scatterer produces a small component $\mathscr{E}_x$ and the evanescent wave is partially converted into diverging one.
The corresponding Pointing vector
\begin{equation}\label{Pointing}
\vec{S}=\frac{c}{8\pi}\Re\left(
   \mbox{\mathversion{bold}$\mathscr{E}$}
     \times
    \mbox{\mathversion{bold}$\mathscr{H}$}^*\right)
\end{equation}
acquires a nonzero $y$-component, $S_y$, so the energy flux outgoing from the plane interface appears. The flux of scattered
wave is calculated at distance $\sim2\lambda$, i.e. in the wave zone, and
normalized by the average flux of the incoming wave in the first medium $S_{\rm
inc}=c\mathscr{H}_{\rm inc}^2/8\pi\sqrt{\eps_1}$. Indicatrix of the scattering
into the upper half-space is shown in Fig.~\ref{fig:double}. Hereafter the
insets show the configuration of scattering bodies. The scattering is minimal in
normal or longitudinal direction and maximal at some medium angles. It is
the quadruple contribution due to the field of image source (\ref{image}) and decay of the evanescent wave amplitude with $y$. The
angle $\varphi$ of the first maximum increases with $\eps_1$. For $\eps_1=2$
the first maximum is $23^\circ$ and for $\eps_1=3$ it is $41^\circ$.  The
forward-backward asymmetry of indicatrix is a clear evidence of violation of the dipole approximation owing to finite sizes ($2kR\approx0.8$).

\begin{figure}\centering
\psfrag{x1}{\large $x, \mu$m}
\psfrag{H}[tl][tr]
{$\mathscr{H}/\mathscr{H}_{\rm inc}$}
\includegraphics[width=0.9\columnwidth]{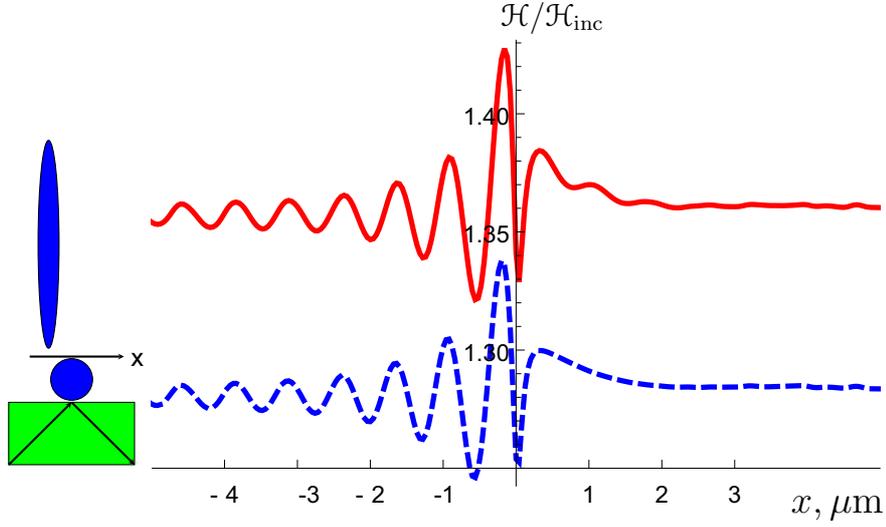}
\caption{The magnetic field at the bottom point of ellipse as a function of
coordinate $x$ at distance $y=0.15$ from the plane  (solid line). Parameters
are $\lambda=1.512~\mu$m, $R=0.06~\mu$m, $\delta=0.01~\mu$m,
$\eps_1=\eps_3=\eps_4=2.25, \eps_2=1, \theta_1=\pi/4$. The same without the
ellipse ($\eps_4=1$, dashed).}\label{fig:line}
\end{figure}

The method can be applied also to different cylinders.  We choose the first cylinder ($\eps_{\textrm{in}}=\eps_3$) with round cross section whereas that of the second cylinder ($\eps_{\textrm{in}}=\eps_4$) is a prolate
ellipse with minor semiaxis $a=0.04~\mu$m and axis ratio $b/a=10$. The major
axis is chosen along $y$-direction. Fig.~\ref{fig:line} shows the field near
the tip as a function of coordinate $x$ at $y=0.15~\mu$m, when the ellipse is
moving along to $x$ axis. The pattern in the near-field domain is much more complicated than in the wave zone. We see  the slight low-frequency oscillations at the right side and deep high-frequency at the left. The physical nature of the
oscillations consists in the interference between falling evanescent wave and
the diverging wave scattered by the circle. They are counter-propagated at the
left and co-propagated at the right. Their spatial frequencies are different
$k_\pm=k_x\pm k_2$. However, in considered case $k_x=k_1\sin\theta_1\approx
k_2$, then the interference oscillation at the right side has a nearly zero
frequency. The slight interference pattern at the right in this case is caused
by the scattering by the tip and these oscillations
vanish without the tip.

\begin{figure}\centering
\psfrag{Abs}[tr][Bl]{\large $x, \mu$m}\psfrag{Ord}[tl][Bl]{Flux}
\includegraphics[width=0.8\columnwidth]{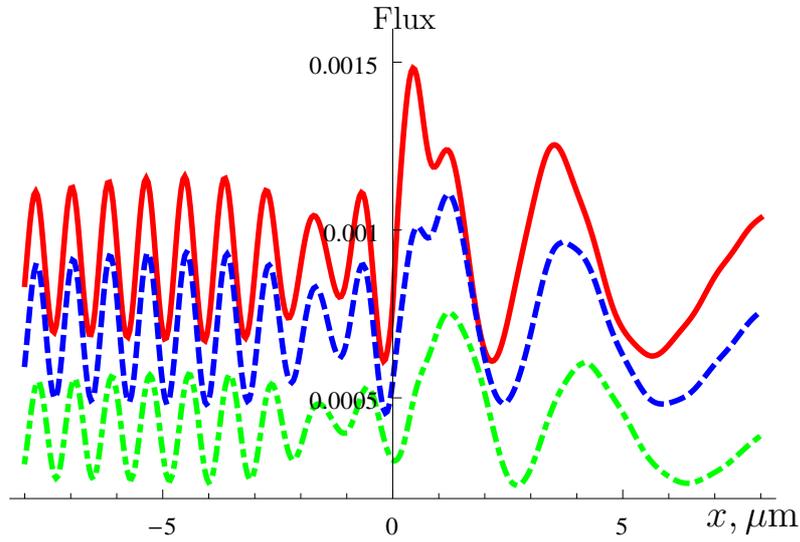}
\caption{Energy going through the central cross section in $y$-direction of the
ellipse as a function of its horizontal position $x$ at distance from the plane
$y=0.15$ (solid line), 0.35 (dashed), 0.50~$\mu$m (dot-dash). Parameters are
$\lambda=1.512, R=0.06, \delta=0.01~\mu$m,
$\eps_1=\eps_3=\eps_4=2.25, \eps_2=1, \theta_1=\pi/4$.}\label{fig:elliptic}
\end{figure}

Fig.~\ref{fig:elliptic} shows the energy going along the major axis through the ellipse middle cross-section as a function of coordinate $x$, namely, the component $S_y$ of vector (\ref{Pointing}) averaged over the horizontal cross-section. The
vertical distance of the tip from the interface is $d=0.15\div0.50~\mu$m, then the minimal distance between the tip of ellipse and the circle starts from
$0.02~\mu$m. We see interference oscillations in the coordinate dependence,
the amplitude decreasing with the distance between the tip and the plane. As follows from Fourier transform (\ref{Green-Fourier}), the higher spatial harmonics with
$q^2>k_2^2$ decay exponentially with distance like $\exp(-\mu_2y)$. Then the
small details are not visible in the far-field pattern. However at $ky\ll1$ the
exponent is not negligible, and then the small-scale details become apparent.
The near field can be observed if one extract the signal and transfer it to the
far zone. In our calculation the stretched ellipse plays a role of such a transmitter.

The energy flux through the far central plane is a simplest (two-dimensional in our case) model of NSOM \cite{Novotny_book,Greffet97,Hecht00}. The curves in
Fig.~\ref{fig:elliptic} correspond to instrumental function of the microscope.
Although, this statement should not be taken literally. The multiple scattering leads to back influence of the probe to the object then it is not a usual linear function
of response.

The coupled boundary equations describing the scattering of the
evanescent wave are solved for two cylinders. The asymmetry of indicatrix and oscillations in the coordinate dependence are observed. The  BEM with proposed Green's function is rather general and applicable for any contour $\Gamma$ or several contours. It can be extended also for 3D geometry. The Green function
could find applications in other calculations, e.g. volume integral equations
including the Born series and discrete dipole approximation.

Authors are grateful to E. V. Podivilov for helpful discussions. This work is
supported by the Government program NSh-4339.2010.2, program \# 21 of the
Russian Academy of Sciences Presidium, and interdisciplinary grant \#42 from
the Siberian Branch of RAS.

%

\end{document}